\documentclass[trackchanges]{aastex701}

\usepackage{graphicx} % Required for inserting images
\usepackage{amsmath,amssymb}
\usepackage{caption}
\usepackage{subcaption}
\usepackage{xcolor}
\usepackage{subcaption}

\newcommand\bowls{\emph{H$_2$O-types}}
\newcommand\bowl{\emph{H$_2$O-type}}
\newcommand\cliffs{\emph{organics-types}}
\newcommand\cliff{\emph{organics-type}}
\newcommand\dds{\emph{CO$_2$-types}}
\newcommand\dd{\emph{CO$_2$-type}}
\newcommand\org{\emph{organic-rich}}
\newcommand{\edit}[1]{{\color{black}#1}}

\begin{document}

\title{Linear Continuum Modelling to Explain The Majority of Bulk Features of Kuiper Belt Object Spectra}
\date{July 2025}

\author[orcid=0000-0001-6680-6558,sname='Fraser']{Wesley C. Fraser}
\affiliation{National Research Council, Herzberg Astronomy and Astrophysics Research Centre, 5071 West Saanich Road, Victoria, BC}
\affiliation{Department of Physics and Astronomy, University of Victoria, Elliott Building, 3800 Finnerty Road, Victoria, BC V8P 5C2, Canada }
\email[show]{wesley.fraser@nrc-cnrc.gc.ca}  

\author[orcid=0000-0002-8032-4528,sname='Buchanan']{Laura E. Buchanan}
\affiliation{Department of Physics and Astronomy, University of Victoria, Elliott Building, 3800 Finnerty Road, Victoria, BC V8P 5C2, Canada }
\email{lbuchanan@uvic.ca}

\author[orcid=0000-0001-9665-8429,sname='Wong']{Ian Wong}
\affiliation{Space Telescope Science Institute, 3700 San Martin Drive, Baltimore, MD, 21218, USA}
\email{ian.wong@nasa.gov}

\author[orcid=0000-0002-6117-0164,sname='Holler']{Bryan Holler}
\affiliation{Space Telescope Science Institute, 3700 San Martin Drive, Baltimore, MD, 21218, USA}
\email{bholler@stsci.edu}

\author[orcid=0000-0002-8255-0545,sname='Brown']{Michael E Brown}
\affiliation{Division of Geological and Planetary Sciences, California Institute of Technology, Pasadena, CA, 91125, USA}
\email{mbrown@caltech.edu}

\begin{abstract}

The first analyses of the James Webb Space Telescope spectra of trans-Neptunian Objects (TNOs) revealed three discrete types of surfaces. This seems to contradict ground-based spectro-photometric datasets, which suggest a continuum of colors with only two surface types.  Here we present linear spectral modelling that reconciles these two results. In our model, the sole parameter is the object's optical slope, and the reflectance spectrum at all wavelengths is linearly proportional to that color, with the slope of that function evaluated from the spectra themselves. When applied to small ($H>5$ for \bowls~ and $H>4$ for the merged sample of \emph{organics} and \dds) and distant ($q>18$~au) objects, we find that this model does a reasonable job of reproducing the overall spectral behaviour of both samples.  Bootstrapping simulations show that if the optical slope were not a good predictor of an object's spectrum, then finding an explained variance of the model that is better than observed occurred in 2.3\% of realizations for the \bowls~ and 0\% of realizations for the \org~ sample. In a $\chi^2$ sense, the optical color is a better predictor of most spectra as compared to the mean spectrum of a class. The trends of optical color and spectra band-areas exhibited for many key compositional materials are well reproduced, and demonstrate that those materials govern the overall spectral shape within a class. Importantly, these results require that within a given class, the band-areas of those key materials are predictable given only its optical color and its surface type.  Unsurprisingly, our simple one-parameter model does not account for the full spectral diversity of TNOs. We speculate that albedo encapsulates much of the remaining diversity.
\end{abstract}

\keywords{\uat{Trans-Neptunian Objects}{1705} --- \uat{Observational Astronomy}{1145}}

\section{Introduction}

Before the on-rush of early spectra from the James Webb Space Telescope (JWST), compositional studies of Kuiper Belt Objects (KBOs), or Trans-Neptunian Objects (TNOs) were mostly restricted to ground-based spectroscopy typically for wavelengths shorter than $\lambda\lesssim3\,\micron$, and to the $\sim50$ known objects brighter than $V\sim23$ \citep[for a recent review, see][]{Barucci2020}. The bulk of TNOs were fainter than this practical depth, limiting their observational studies to spectro-photometric observations. 

The optical color surveys demonstrated that for small TNOs, fainter than an absolute magnitude $H\sim6$, the color distributions presented at least two distinct groups of objects \citep[e.g][]{Barucci2004,Peixinho2012,DalleOre2013,Schwamb2019} that have different mean color distributions. The addition of NIR color observations revealed that generally the bulk of TNOs fall into only two classes that exhibit a range of correlated optical-NIR colors \citep{Fraser2012,Fraser2023,Bernardinelli2025}.

The last few years of observations with JWST have revealed a wealth of specific compositional information of TNOs, the breadth of which the community is still coming to understand. \citet{Pinilla-Alonso2025} report that beyond the dwarf planets which appear to show a broad range of rheologies, smaller TNOs tend to fall into three distinct categories. Following the naming convention of \citet{Holler2025}, the \bowls~ (formerly known as bowls) whose reflectance spectra are dominated by water-ice; the \cliffs~ (formerly known as cliffs) who appear to be relatively water-ice poor with clear signatures of aliphatic organics and CO$_2$ ice, and the equally water-ice poor and CO$_2$ rich \dds~ (formerly known as double-dips) which show clear signatures of CO absorption \citep{DePra2025}. Considering specific compositional variations, the three-group taxonomy can be further subdivided \citep{Brunetto2025,Holler2025,Wong2025b}.

\begin{figure}[ht!]
\begin{center}
\includegraphics[width=0.8\textwidth]{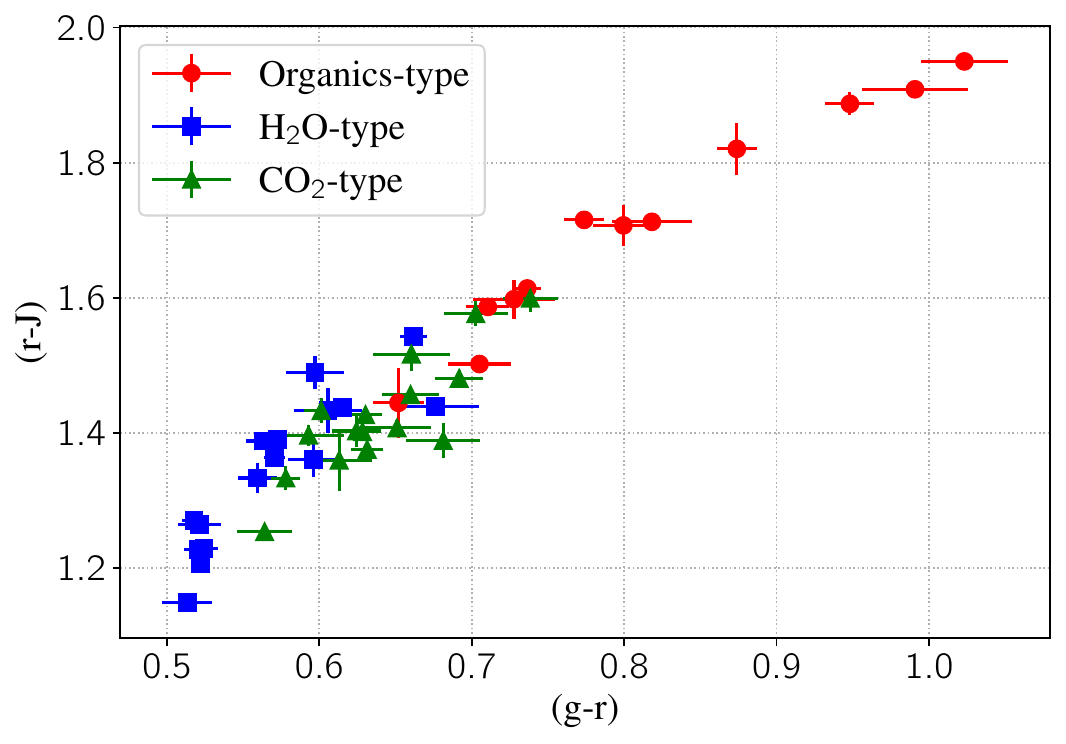}
\caption{Estimated intrinsic (g-r) and (r-J) colors of NIRSpec targets, showing \bowls~ with $H>5$ and $q>18$~au and $H>4$ and $q>18$~au for \dds~ and \cliffs. Reflectance at $0.477\, \micron$ (g) and $0.62\, \micron$ (r) were extrapolated from a linear fit the of the spectrum at longer wavelengths, and the reflectance at $1.22\mbox{ $\mu$m}$ (J) was drawn from the spectra themselves. Uncertainties were estimated by randomly scattering the spectra in the fitted region according to the uncertainties in the spectra, refitting, and repeating that process 25 times. Points are labelled according to the nomenclature introduced by \citet{Holler2025}. \label{fig:colors}}
\label{default}
\end{center}
\end{figure}

The conclusion that small TNOs exhibit three broad surface types with further detailed subdivisions therein seems in apparent conflict with the results of ground-based spectro-photometry which suggest only two classes each with a continuum of surface colors. The classification from JWST spectra was made from a principal component analysis (PCA) of those spectra. In that analysis, the \bowls~ appear distinct from the \dds~ and \cliffs. Though they have been classed differently based on specific compositional components, \citet{Pinilla-Alonso2025} and \citet{Brunetto2025} points out that the \cliffs~ and \dds~ tend to fall along a continuum in principal component space. Indeed, as shown in Figure~\ref{fig:colors}, the (g-r) and (r-J) colors estimated from the spectra appear to exhibit a continuum, suggesting that plausibly, the JWST spectra do not indicate the presence of three unique classes, but rather may show a smaller number of classes with a continuum of spectra. We address this possibility in this manuscript.

\section{Spectral Modelling}

In the works of \citet{Fraser2023} and \citet{Bernardinelli2025}, the optical-NIR colors of TNOs fall into the BrightIR and FaintIR classes, which \citet{Bernardinelli2025} refer to as NIRB and NIRF, respectively. Each class shows a similar broad range of optical colors that strongly correlate with reflectance beyond $\lambda\sim0.9\,\micron$, but the FaintIR/NIRF class has lower NIR reflectance for a given optical color. \citet{Fraser2012} and \citet{Fraser2023} suggest that the range of colors of each class can be accounted for by a simple linear mixing model where each class is defined by a pair of end-member spectra, and the specific spectrum of a given object is then just a specific geometric mixture of those two end members. In this way, the correlated optical-NIR colors of a class can be reproduced. From this model, it follows that given an object's optical spectral slope $o$, and its compositional class, its NIR reflectance can be accurately predicted.  \edit{If it holds true that a TNO's optical color and its NIR reflectance spectrum are correlated, then it follows that optical color will be a good predictor of not only a TNO's spectral class, but also its composition. Such a result would add great value to upcoming large surveys, such as the Legacy Survey of Space and Time, or the Roman Community Surveys, which will provide tens of thousands of new objects with accurate broad-band colors which would then provide reliable spectral classifications. Such a finding would also rectify the apparent conflict between spectral classification from JWST and the wealth of broad-band spectro-photometry}. The question remains, is this result lost when significantly higher resolution and bandwidth spectra are considered.

A prediction of the linear modelling approach of \citet{Fraser2012} is that the reflectance $R$ at some wavelength $\lambda$ can be modelled as

\begin{equation}
R\left(\lambda\right) = m\left(\lambda\right)o + b\left(\lambda\right).
\label{eq:model}
\end{equation}

\noindent
$\lambda$-dependent parameters $m\left(\lambda\right)$ and $b\left(\lambda\right)$ are just the two components of the linear model. That is to say, with knowledge of the spectral slope $o$, the entire spectrum of an object can be recreated.

To test the validity of this prediction, we make use of published JWST NIRSpec observations. The bulk of the data we make use of comes from the Cycle 1 program, \emph{Discovering the Surface Composition of trans-Neptunian objects} \citep[DISCOs GO-2418,][]{Pinilla-Alonso2025}. We also make use of the spectra of Neptune Trojans \citep[GO-2550,][]{Markwardt2025} and of a few distant TNOs observed in program GO-4665 (PI Holler, B.). All reductions followed the methods of \citet{Wong2024}, \edit{which provides calibrated spectra and spectral variances for each NIRCam spectral channel}. 

We were unable to find reliable optical spectral slopes from the literature; many objects were missing published values, and inconsistent values were found where repeat measurements were published, possibly as a result of filter transformations or inconsistent calibrations. Instead we make use of linear fits to the \edit{shorter wavelengths in each of the available} NIRSpec spectra to estimate spectral slopes. Spectral slopes, $o$ were fit to each spectrum using linear regression weighted with the variance of each spectral \edit{channel} to wavelengths shorter than some value $\lambda_o$. We adopt $\lambda_o=1.2\,\micron$ for \bowls\, and $\lambda_o=1\,\micron$ for all other types. 
The value of $\lambda_o$ was selected as where visually the majority of objects in our sample exhibit nearly linear spectra. Some objects exhibit slight curvature at the longest wavelength end of this range, though we adopt this value as a compromise between selecting the most linear region, and enough signal for a meaningful fit. There is some sensitivity in the results from the choice in $\lambda_o$, in terms of $\chi^2$ values of the model, but the statistical significance of the improvement remains strong. To that point, we also discuss $\lambda_o=1.0\,\micron$ for \bowls~  and $\lambda_o=0.85\,\micron$ for the \org~ objects which guarantees only linear regions of the spectra are used in fitting $o$ for those objects.

To select spectra of small TNOs, absolute magnitudes were collected from a variety of sources. We started with  $H_r$ values published in \citet{Peixinho2015}. Next, values from the MBOSS2 dataset \citep{Hainaut2012} were used. Next, values from \citet{Sheppard2012} and \citet{Benecchi2013} were used if not available in \citet{Peixinho2015} or \citet{Hainaut2012}. Finally, missing values were drawn from JPL Horizons. These latter values may be considered less reliable, and are in V-band. As mentioned, we were unable to find reliable optical colors. As such, we chose to leave the $H$ values in their reported bands resulting in values of $H$ that are somewhat inhomogenous in their band and quality. Though this collection of $H$-values is not used for anything more than sample restrictions.

We restrict our consideration to objects with perihelia $q>18$~au to avoid any possible surface evolution driven by cometary activity \citep{Jewitt2009} which the majority of more distant bodies would not undergo. We also  \edit{apply a restriction in H, considering objects smaller than some value}. This restriction is driven by the appearance of the two different classes, for objects with $H\gtrsim6$ in the spectro-photometry \citep{Fraser2012,Peixinho2012}. Given the available spectra, limiting the sample to objects with $H>6$ is too restrictive for any meaningful modelling analysis. Rather, we examine various limits on $H$ and note that modelling of the $H>6$ range will require more data for smaller TNOs. We also avoid the known Haumea family members in the sample, as these are objects with a unique collisional origin and resultant unique spectra \citep{Brown2007,Pinilla-Alonso2025} that are not shared by the bulk Kuiper Belt.

Fits of $m\left(\lambda\right)$ and $b\left(\lambda\right)$ were done in a least-squares sense making use of the spectral uncertainties. 
To fit the linear spectral model of Equation~\ref{eq:model} to a given sample (e.g. the \bowls), all spectra were first normalized to unity at $1\,\micron$. This choice in wavelength was somewhat arbitrary, and was chosen as a clean region of high signal to noise (SNR) that was uninfluenced by any obvious absorption features. The mean spectrum $R_{\textrm{mean}}$ of the sample was then subtracted from individual spectra. \edit{Then, at each wavelength the residual spectra of the sample and the sample's} optical slopes $o$ were then fit with the model of Equation~\ref{eq:model}   in a least-squares sense while respecting the variance values \edit{ of each object at that wavelength}\footnote{\edit{We point out that with gaussian distributed measurement uncertainties, the use of least-squares and maximum likelihood formulations are fully equivalent. As well, the use of a more complex minimization routines such as Markov-Chain Monte Carlo techniques are unnecessary as they provide the same information as the best-fit $(m, \, b)$ and covariance matrix that the linear regression provides.}}.  With the $m\left(\lambda\right)$ and $b\left(\lambda\right)$ (which for a given wavelength are just constant values), the spectrum \edit{ of a given object}, can then be produced from  its optical slope $o$ by reversing these steps. That is, $R_{\textrm{model}}\left(\lambda\right) = R_{\textrm{mean}}\left(\lambda\right) + m\left(\lambda\right)o + b\left(\lambda\right)$.

\begin{figure}[h!]
\centering
  \includegraphics[width=.95\linewidth]{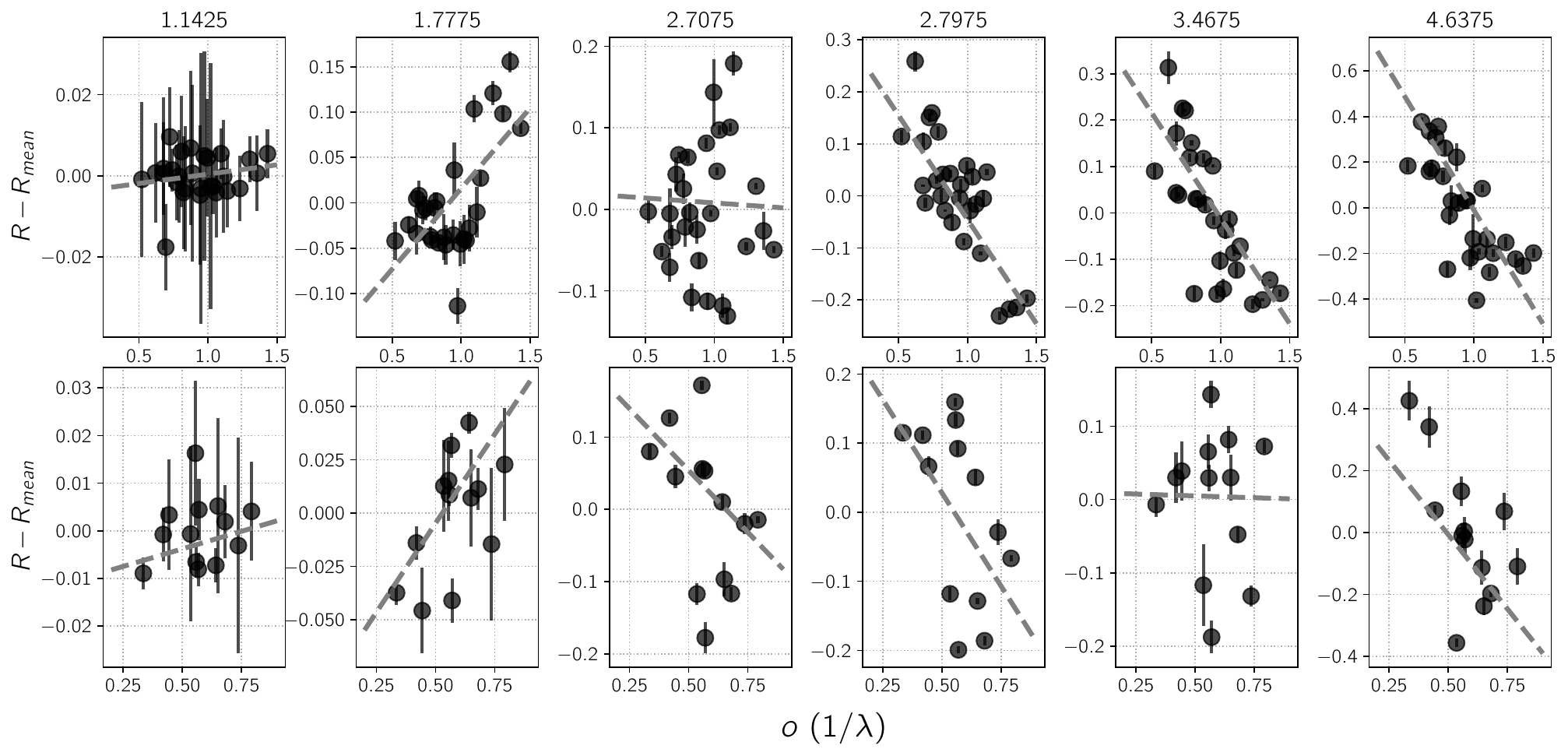}
\caption{Demonstrations of the fits our linear model at five different wavelengths (labels are in microns). The top row shows  the \org~ sample, which includes \cliffs~ and \dds~ with $q>18$~au and $H>4$. The bottom row shows the \bowls~ with $q>18$~au and $H>5$ (see Section~3 for further discussion). Plotted is reflectivity minus mean reflectivity vs optical slope $o$. The grey lines shows the linear model fit at each wavelength.}
\label{fig:fits}
\end{figure}

\begin{figure}[ht!]
\begin{center}
\includegraphics[width=0.7\textwidth]{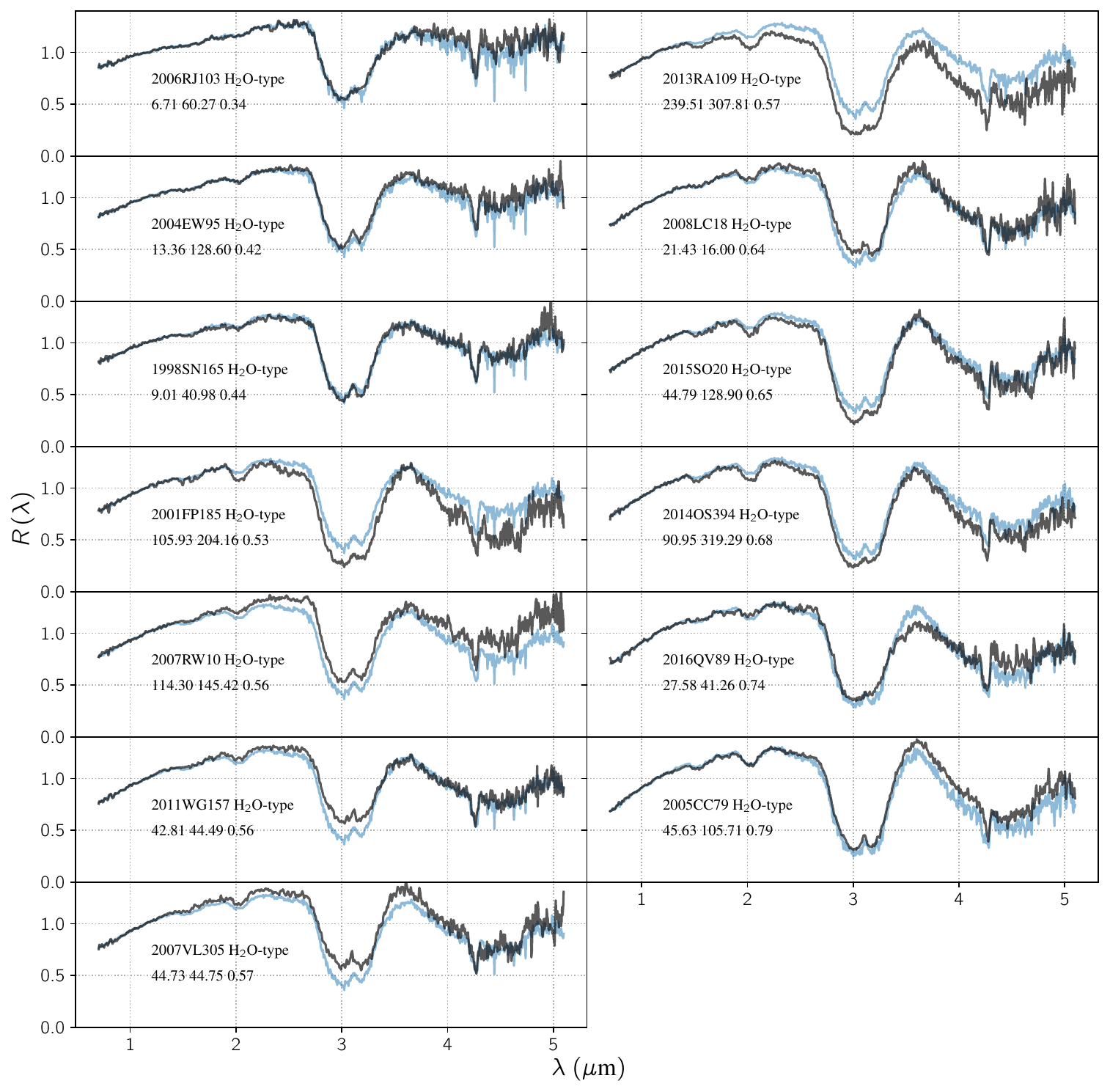}
\caption{Observed (black) and model (blue) spectra of the 13 \bowls~ that satisfy our $q$ and $H$ criterion. For each spectrum, the designation of the object is shown. The second line of text in each panel lists the reduced $\chi^2$ using the linear model, the reduced, $\chi_{mean}^2$ when considering the mean \bowl~ spectrum, and the value of the optical spectral slope, $o$, for that object. The SNR and small sample manifests as scatter in the model which can be seen especially at the longest wavelengths where the SNR of the sample is lowest. A larger sample would be useful in supressing this effect.  \label{fig:bowls}}
\label{default}
\end{center}
\end{figure}

\begin{figure}[h!]
\begin{center}
\includegraphics[width=0.7\textwidth]{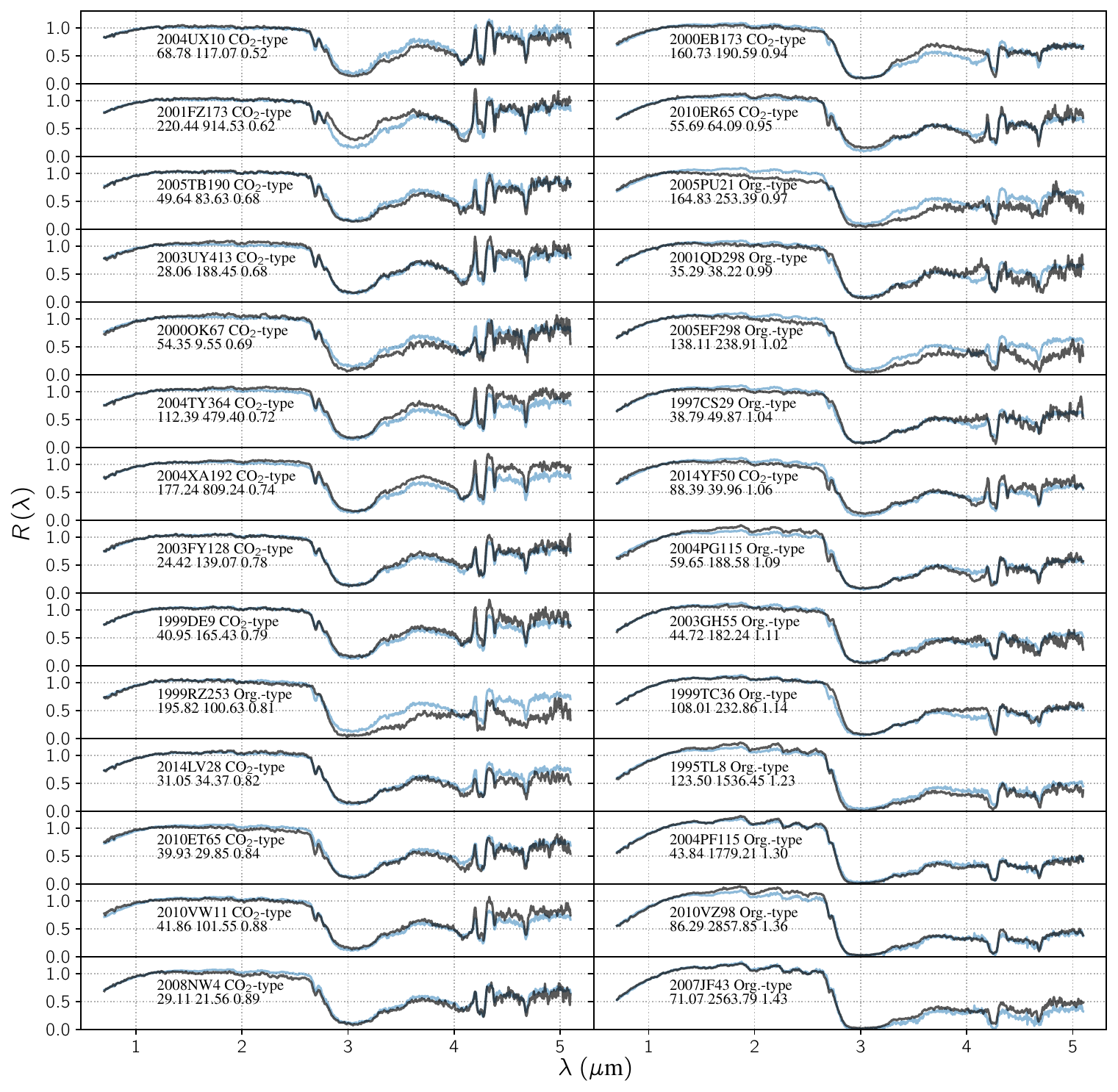}
\caption{Observed (black) and model (blue) spectra of the 28 \org~ types that satisfy our $q$ and $H$ criterion. For each spectrum, the designation is shown. The second line of text in each panel lists the reduced $\chi^2$ using the linear model, the reduced $\chi_{mean}^2$ when considering the mean \org~ spectrum, and the value of $o$ for that object. \label{fig:tts}}
\label{default}
\end{center}
\end{figure}

To demonstrate the model fitting, we show evaluations of the vectors of the linear model $m(\lambda)$ and $b(\lambda)$ at five different wavelengths in Figure~\ref{fig:fits}. The fit is a least squares linear regression of the optical spectral slope, $o$, and reflectivity minus mean reflectivity of the sample at a given wavelength, with data points weighted as the inverse square-root of their uncertainties. Most panels show well behaved regions of the spectra where SNR at that wavelength is similar for each object. Problematic wavelengths are shown for the organic-types at $\lambda=2.7075\mbox{ $\mu$m}$ and $\lambda=3.4675$ for the \bowls. These wavelengths show negligible correlation with optical slope, and so the model accounts for none of the scatter there. The influence of objects with particularly high SNR compared to the sample can be seen with fits overly weighted towards those objects (e.g., the offset in fit at $\lambda=4.6375$ for the \bowls). That circumstance  is the majority source of the model noise that can be seen in Figures~\ref{fig:bowls} and \ref{fig:tts}, which is most prominent at longer wavelengths. In an attempt to suppress this model noise, we explored alternative fitting procedures, including rejection of the most discrepant data point at each wavelength, or weighting by different powers of the uncertainty, but found that when fits were chosen such that model noise was minimized, it tended to overly favour the few objects with the highest signal to noise, and so be a poor representation of the whole sample. A better solution is to acquire more observations with a higher overall SNR for the fainter sources in the sample. 

\begin{figure}[ht!]
\begin{center}
\includegraphics[width=0.6\textwidth]{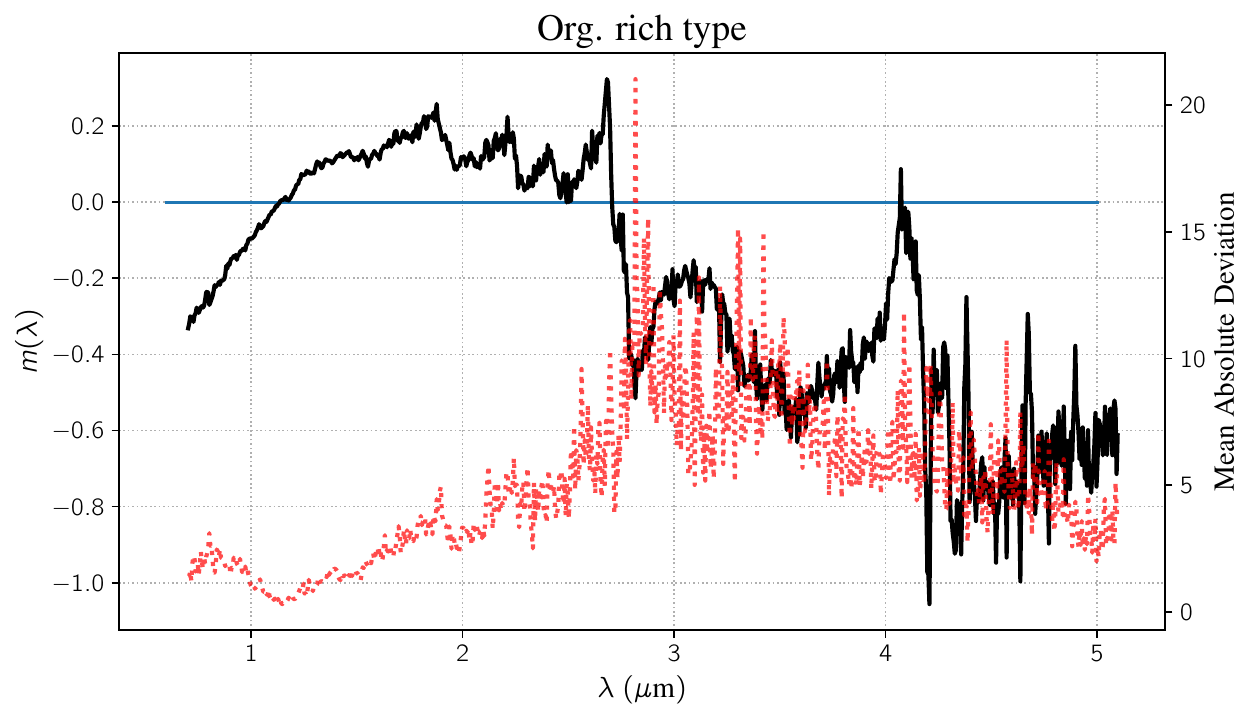}\\
\includegraphics[width=0.6\textwidth]{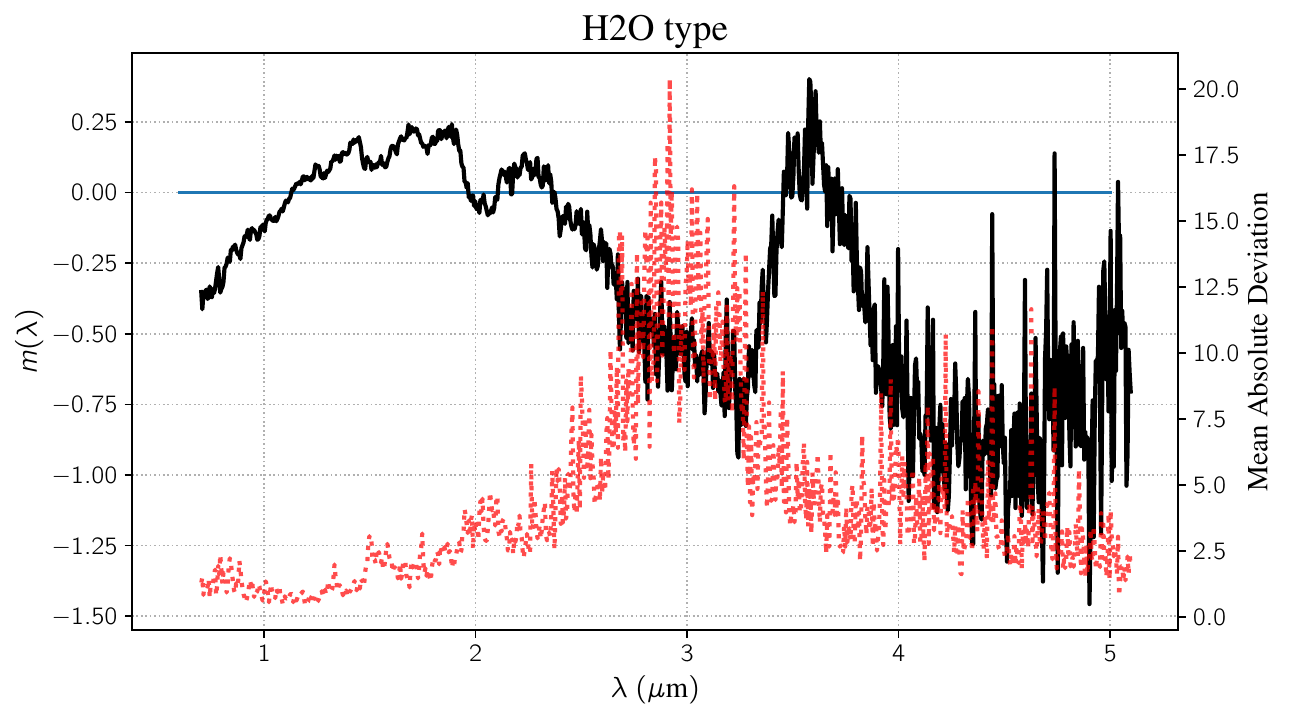}
\caption{ {\color{black} Results of the linear fits of the \org~ and H$_2$O-rich samples showing $m(\lambda)$ as a function of wavelength (black). Where the curve is positive, the model spectra will increase with increasing spectral slope, and where negative will decrease with $o$. The hatched red curve shows the SNR-weighted mean absolute deviation (MAD) between the model reflectance and linear fits  (e.g., in Figure~\ref{fig:fits}). Some spikes in MAD correspond with regions where the linear model fails to account for the full spectral diversity of the sample, such as at the $2.7\micron$ CO$_2$ feature, and the jump in reflectivity associated with the $4.2~\micron$ CO$_2$ feature. Other spikes in MAD correspond to regions dominated by problematic wavelengths where fits are overly weighted to objects of particularly high SNR compared to the rest of the sample. }  }
\label{fig:correlation}
\end{center}
\end{figure}

The fitted $m\left(\lambda\right)$ for the \org~ and H$_2$O types are shown in Figure~\ref{fig:correlation} which shows the spectral regions that are most strongly and most weakly correlated with optical slope.  As well, that figure shows regions that exhibit the highest model noise, as evaluated by the mean absolute deviation of the model residuals. The  absorption bands common in TNO reflectance spectra feature prominently in $m\left(\lambda\right)$, including absorptions by OH, methanol, CO, CO$_2$, and H$_2$O.

The model described by Equation~1 is similar to principal Component Analysis  \citep[PCA,][]{Pinilla-Alonso2025} in that both are linear models, whereby a model spectrum is simply the sum of one or more linearly scaled  basis  vectors (one here, two in the PCA analysis of \citet{Pinilla-Alonso2025}). The main difference between the two models is that the PCA analysis is unconstrained, and utilizes as a base the vector that minimizes residual variance between model and data, while the model we present has the constraint that the model spectrum of a source is proportional to that source's optical slope. We facilitate comparison of the two models by applying 1 dimensional PCA to each of the samples we model. As we discuss below, both our linear model and the  1 dimension PCA have comparable performance in matching observations. Expectedly, the restriction of proportionality with optical slope results in a higher residual variance for our linear model than does the PCA technique. The utility of this approach however, is to directly connect optical color and spectral properties across a broad range of wavelengths. More fundamentally, PCA provides little predictive power. Meanwhile our approach allows for a prediction of a newly discovered object's spectrum, with just a measure of its optical color, measurements that will be available in abundance from the Legacy Survey of Space and Time \citep{Kurlander2025}.

\section{Modelling Results}

We first consider ~\bowls~ with $H>5$, resulting in a sample of 13 objects. In Figure~\ref{fig:bowls} we present the observed  and linear modelled spectra, $R_{\textrm{model}}\left(\lambda\right)$.  As can be seen, the model does a reasonable job of reproducing the spectral behaviour across the entire coverage. Notably, the match shortward of $\sim2\,\micron$ is excellent for most objects. Importantly, the depths of the water-ice, and $CO_2$ features are well reproduced, including the presence of the $3.1\,\micron$ water-ice fresnel feature where it is present in the observed spectra, and absent when it is not. The deviations of model from spectrum, \edit{ which can be seen in Figure~\ref{fig:bowls}}, appear to be mostly systematic, whereby the residual between model and observed spectra seem to show a broad spectral slope. We speculate that a NIR slope as a second fitting parameter \edit{might} bring the discrepant models in much tighter agreement with the data.

To assess how close the models are to the data, we report reduced $\chi^2$ values. As a guide of how much of the spectral behaviour is accounted for by our model, we compare $\chi^2$ to the reduced $\chi_{mean}^2$ values between observation and the mean \bowl~ spectrum. When calculating $\chi^2$, we consider only data for $\lambda>\lambda_o$, avoiding the region used in measuring $o$. For 12 of 13  \bowl~spectra, the model is a better match to the observation than is the mean, demonstrating that the spectra exhibit an evolution with optical slope $o$ which is encoded in the linear spectral model. We emphasize that the models in Figure~\ref{fig:bowls} are not fits to individual spectra, but rather just a reproduction given the spectral slope for $\lambda<\lambda_o$, and the model of Equation~\ref{eq:model}. The $\chi^2$ values can be lowered further by allowing $o$ to vary as a free parameter. It may be that high-quality measurements of the optical slope allow for better spectral reproductions than do the extrapolated $o$ values we make use of here.

We make use of the reduction of $\chi^2$ compared to $\chi_{mean}^2$ as a means to test the significance of representation of the observations by the model. We apply the same modelling analysis from above with bootstrapped values of $o$. After randomizing the $o$ values through bootstrapping by reassigning observed values amongst the sample, we refit the spectra with the linear model, and count the number of spectra with $\chi^2$ that are lower than $\chi^2_{mean}$. For \bowls~, 0.2\% of 2,000 randomized modelling iterations resulted in 12 or more of the 13 spectra with  $\chi^2<\chi^2_{mean}$. When considering $\lambda_o=1.0\,\micron$, 11 of 13 \bowls~ have $\chi^2<\chi^2_{mean}$, which occurred 45 times in 1,000 iterations of the randomized $o$ simulations. We speculate that this reduction in significance is not a result of the model, but instead due to the quite limited range of data from which to measure $o$, and the low SNR of the spectra in that range.  (see Figure~\ref{fig:bowls}).

In a similar fashion, we make use of the explained variance, which is the fractional variance of the sample explained by the model.  The linear model has an explained variance of 
75\% for the \bowls. When evaluating the explained variance of the samples with bootstrapped values of $o$ described above, we found that only 2.3\% of the bootstrapped realization of the \bowls~ had model explained variances better than the observed value. 

We also considered including larger \bowls~ in the modelled sample. Decreasing the absolute magnitude limit to $H>4$ includes an extra 4 objects. The modelling results however, are significantly worse; the number of objects with $\chi^2 < \chi^2_{mean}$ is 11 of 16. The four larger objects include objects like 2002 XV93 and 2004 NT33, both of which show significantly deeper water-ice bands than do similarly colored but smaller sized objects. This is consistent with the findings of \citet{Brown2012} and more recently \citet{Wong2025a} who showed that larger \bowls~ contain higher concentrations of surface water-ice, and demonstrates that this linear modelling is only successful for the sample of smaller-sized ~\bowls. This is also broadly consistent with spectral photometry that shows a transition in the color distributions that separates mid-sized TNOs from smaller objects, though the NIRCam spectral sample includes too few objects to probe objects with $H\sim6$.

Driven by the observation that the colors of the \dds~ and \cliffs~ tend to broadly fall along a continuum  with \cliffs~ tending to be redder than \dds~ \citep[e.g.][and Figure~\ref{fig:colors}]{Pinilla-Alonso2025} we consider objects with $H>4$ from both types together, which henceforth we label as the \org~ sample for ease of discussion. Their spectra, and the spectra from the linear model, $R_{\textrm{model}}\left(\lambda\right)$, are shown in Figure~\ref{fig:tts}.

Like with the \bowls, many organic rich spectra are well reproduced by the linear spectral model. The linear model has an explained variance of 55\% for the \org~ sample. To test the significance of this result, we applied the same bootstrapping simulations as was done for the \bowls. In those simulations, not a single bootstrapped iteration in 2,000  realizations had a higher explained variance than observed. The reduced $\chi^2$ of the model is better than that from the mean spectrum in 23 of 28 cases. This occurred in only in 0.55\% (11) of 3,000  iterations of the randomized $o$ simulations. When considering a shorter value, $\lambda_o=0.85\,\micron$, the number of objects with lower $\chi^2$ than from the mean reduces to 19 of 28, with an decrease in significance of 6\% of 2,000 randomized iterations. Like when reducing $\lambda_o$ when fitting the ~\bowls, we suspect that this reduction in significance is due to the limited range of data from which to measure $o$, and the low SNR of the spectra in that range. 

Much like the \bowls, the most significant deviations can be roughly described as a slope in the model residual, with the difference becoming most apparent longward of $\lambda\sim2\,\micron$. There are also some notable differences in the shapes of the CO$_2$ and methanol bands. The model consistently  under-predicts the depth of the $2.7\,\micron$ CO$_2$ features for the redder half of the sample. Similarly, for some of the redder objects, the model produces deeper methanol features than present in the spectra. The model does an excellent job reproducing the OH absorption, the centre of the $4.2\,\micron$ CO$_2$ feature, and does a reasonable job matching the shape and depth of the aliphatic absorptions at $3.4\,\micron$, though the model is particularly discrepant in this region for some objects. That the \cliffs~ and \dds~ share quite similar spectra is well reflected in the model, with the majority of the spectral range of both classes being broadly reproduced by a common mixing model. As highlighted by \citet{Pinilla-Alonso2025}, the \dds~ and \cliffs~ differ in their methanol and CO$_2$ absorptions. 

We considered additional groups beyond the \bowls~ and \org~ samples discussed above. For example we modelled the entire merged sample, which failed spectacularly in that the model failed to reproduce the observed spectra in all cases. We also treated the \cliffs~ and \dds~ as separate groups and found that the model performed only slightly better at least when considering reduced $\chi^2$ as the metric. While the CO$_2$ absorption depths were better reproduced, the majority difference between model and observed spectra was still dominated by the residual slope at longer wavelengths that is exhibited by the \org~ sample. We also considered expanding the \org~ sample to smaller H-values, and found reduced match between model and observation, \edit{with fractionally fewer objects being acceptably matched by the linear model as smaller H-value objects were added}. Like as found for the \bowls, it seems the organic rich objects also show a trend with $H$, consistent with the result that the spectrophotometric sample shows trends of color with $H$. Exploration of these trends is left to a future work.

When compared to application of the PCA technique, the linear model fairs relatively well. 1D PCA applied independently to either the \bowls~ or \org~ samples had an explained variance of 80\% for both samples. As mentioned above, the linear model had an explained variance of 55\% for the \org~ sample, and 75\% for the ~\bowls.  The fraction of PCA model spectra with better $\chi^2$ than that of the mean spectrum increases from 23 of 28 to 25 of 28 for the \org~ types, and  remains at 11 of 13 for the \bowls. The apparent lower performance of the linear model is fully expected, as unlike the PCA which is unconstrained to finding the linear model that maximizes explained variance, our linear model requires proportionality of the model outputs with optical spectral slope, and so would never be expected to perform as well as PCA. The results of our linear model show that much of the spectral diversity of the NIRSpec sample is correlated with optical slope. Equally interesting as the regions that show strong correlation with $o$, are those that do not. We discuss these regions, and the physical interpretation afforded by our model below.

\subsection{Absorption Band Areas}

\begin{figure}[ht!]
\begin{center}
\includegraphics[width=0.6\textwidth]{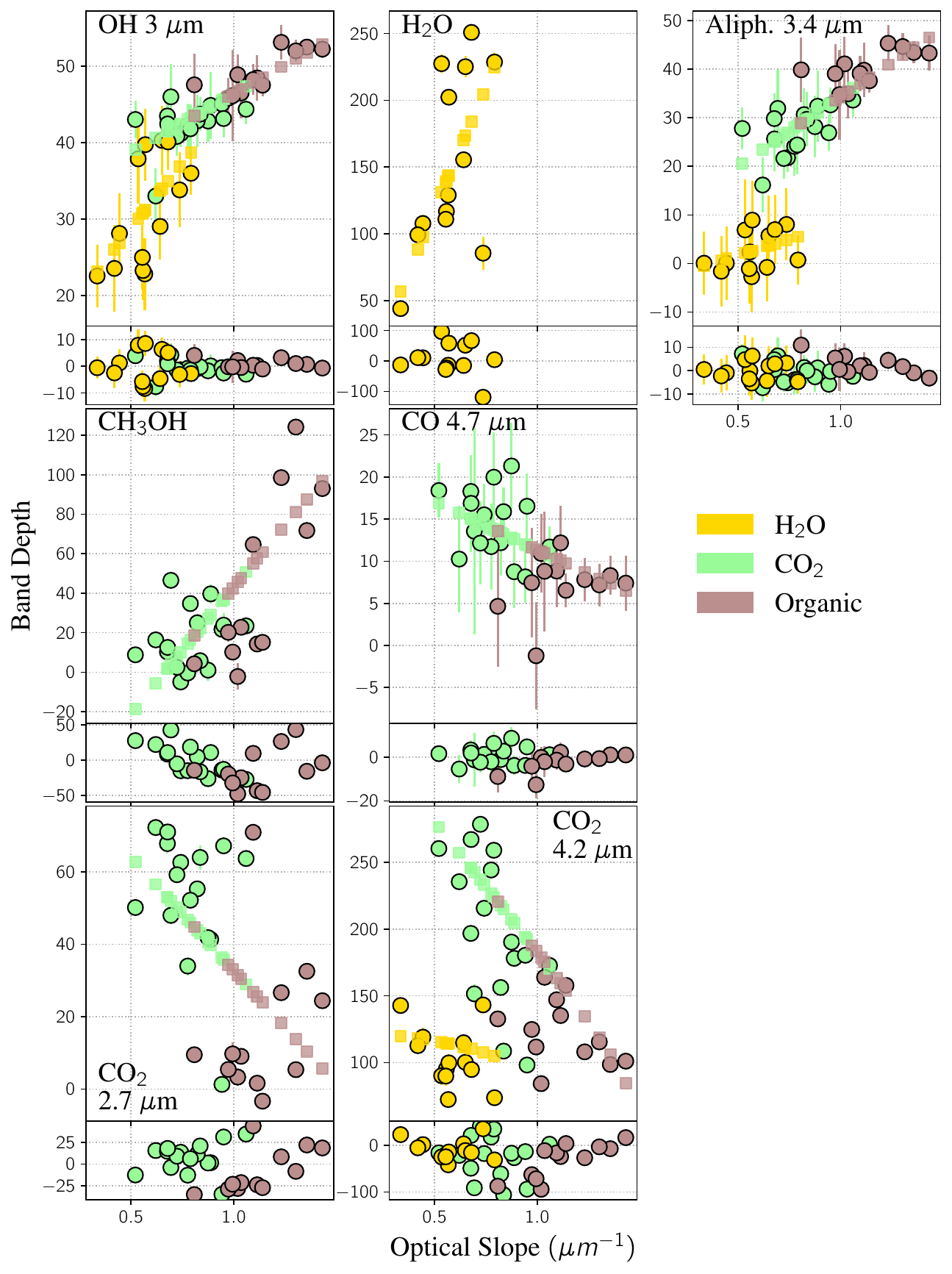}
\caption{Band areas of the prominent H$_2$O ($1.5$ and $2\mbox{ $\mu$m}$ combined), CO$_2$, CO, methanol, and aliphatic absorptions for \bowls~(yellow), \dds~ (green), and \cliffs~ (brown). Units are in reflectivity-angstroms. Values from the observed spectra are marked with circles and from the linear model are marked with squares. Errorbars on the absorption depths are derived from the RMS values of the spectra within each band. The bottom panels show the model residuals. H$_2$O is the sum of band areas of the 1.5 and $2 \,\micron$ water-ice features, and these values are omitted for the \cliffs~ and \dds~ which do not show prominent water-ice absorptions, but instead could be contaminated by methanol. Similarly, we omit the methanol, 2.7 $\micron$ CO$_2$, and $4.7\,\micron$ CO depths for \bowls~ which tend to not show those features. Finally, we note that for \bowls, the depth of the feature in the $3.4~\micron$~region may not be entirely due to aliphatic absorption, but may also include contributions by H$_2$O$_2$ (Protopapa personal communication). \label{fig:depths}}
\label{default}
\end{center}
\end{figure}

In Figures~\ref{fig:depths} we present absorption band areas for the prominent H$_2$O, CO$_2$, CO, methanol, and aliphatic absorptions. We discuss each feature in turn. 

We measure the depth of the  prominent $3\,\micron$ OH feature by \edit{by means of} a linear fit to $1.2<\lambda<2.23\,\micron$, and integrating the difference across $2.9<\lambda<3.2\,\micron$.  This feature shows a correlation of optical slope with absorption depth for both the \bowls~ and \org~ sample. The linear spectral model reproduces this trend. The sample variance of the band areas is reduced from 17.7 to 5.6 and 48.2 to 31 for the \org~ sample and \bowls~ respectively.

For the $H_2$O ice features at 1.5 and 2.0$\,\micron$, continua across the features were evaluated from linear fits to $1.35<\lambda<1.4\,\micron$ and $1.68<\lambda<1.71\,\micron$ and $1.85<\lambda<1.89\,\micron$ and $2.2<\lambda<2.23\,\micron$, with band areas integrated across $1.49<\lambda<1.61\,\micron$ and $1.95<\lambda<2.1\,\micron$, respectively. For the \bowls, the $1.5$ and $2.0\,\micron$ water-ice features also show a clear correlation with $o$, in agreement with the interpretation that the $3\,\micron$ feature is dominated by water-ice absorption. There is a hint of a bimodality in the water-ice band area for the \bowls~ that cannot be accounted for by our simple linear model. The confirmation of this bimodality awaits a larger sample.  In the measured water-ice band-areas for the \bowls~ the sample variance  is reduced by 35\%, from $4.1\times10^3$ to $2.7\times10^3$.

A feature present in the \org~ sample, but not the \bowls~ is the CO$_2$ band near $2.7\,\micron$. For this feature, the continuum was evaluated with a linear fit to the $2.62<\lambda<2.65\,\micron$ and $2.72<\lambda<2.73\,\micron$  regions with band area integrated over $2.68<\lambda<2.71\,\micron$. While the model does reduce the sample variance of the band area,  this feature does not seem to show a band-area correlation with $o$ but rather seems to cluster around two distinct values for the \dds~ and \cliffs, which manifests as a negative correlation of band-area and $o$, driven by the fact that the \cliffs~ tend to be redder and show those features less prominently than the bluer \dds. The $4.2\,\micron$ CO$_2$ absorption band area shows a weak negative correlation with $o$, for both the \bowls~ and stronger so for the \org~ objects, which in both cases is well reproduced by the linear model, though it appears that the trend in the observed spectra is non-linear, which can be explained by the near-saturation of that feature. We omit a presentation of the $2.78\,\micron$ CO$_2$ feature which is not commonly present in either the \bowls~ or \org~ sample.

The CO band-area near $4.7\,\micron$ was evaluated with average continuum from the $4.72<\lambda<4.81\,\micron$ region with band area integrated over $4.66<\lambda<4.7\,\micron$. The band-area  shows a negative correlation with $o$ for the \org~ sample, which is well reproduced by the linear model. The band area sample variance is reduced by 52\%, from $3.3\times10^3$ to $1.6\times10^3$. The negative trend in CO for the \org~ bodies does not appear to be driven by the fact that the \cliffs~ tend to have weaker CO absorption features than do the \dds. Both populations seem to show this trend independently, and indeed this trend appears continuous across the two samples together. 

Other features that are present in the spectra of the \org~ objects and absent in the \bowls~ are:  the aliphatic shelf near $3.4\,\micron$ with a linear continuum fit to $1.2<\lambda<2.23\,\micron$ region that is extrapolated to  $3.3<\lambda<3.6\,\micron$ over which the band area is integrated; and the methanol bands at 2.3 and 2.5$\,\micron$ with continua across the features evaluated from linear fits to $2.21<\lambda<2.23\,\micron$ and $2.39<\lambda<2.41\,\micron$ and $2.39<\lambda<2.41\,\micron$ and $2.55<\lambda<2.58\,\micron$, with band areas integrated across $2.26<\lambda<2.33\,\micron$ and $2.45<\lambda<2.53\,\micron$, respectively. The aliphatic shelf has a weak positive correlation with $o$ which is also well reproduced by the linear model, and results in a significant reduction in band area sample variance, from 58 to 19. The model produces a strong positive correlation of methanol band-depth with $o$  and indeed results in a significant decrease in band area sample variance from $1.0\times10^3$ to $6.0\times10^2$. This trend in the observations is driven by the reddest 5 \cliff~ objects which \citet{Brunetto2025} label \emph{Cliff1 types}. It remains to be seen if this trend in the methanol absorption of the \cliffs~ is a continuum or a bimodality as suggested by \citet{Brunetto2025}. The methanol band area trend appears to be slightly negatively proportional to $o$ for the rest of the \org~ types, suggesting that it is a bimodal behaviour. \edit{This bimodal behaviour has been interpreted as evidence for two subclasses of the organic objects, which \citet{Brunetto2025} labelled the \emph{Cliff-1} and \emph{Cliff-2} types. Additional spectra of the \org~types are needed to determine if this is a trend with color or bimodality in methanol absorption.}

In summary, the spectra of the \bowls~ are generally well reproduced by the linear spectral model which well reproduces the majority of the observed spectra across the most of the NIRSpec range. Trends in absorption band-areas of water-ice, CO, and CO$_2$ with $o$ are well reproduced. For the organic rich bodies, the linear model is similarly successful in a $\chi^2$ sense, and reproduces the trends in OH, CO, aliphatic organic, and $4.2\,\micron$ CO$_2$ absorptions with $o$, but fails when compared against methanol, and shorter wavelength CO$_2$ features. Guided by the very low occurrence rate of the model spectra having $\chi^2<\chi^2_{mean}$ when the values of $o$ are scattered, it appears that the simple linear model we present here does indeed account for the observed trends moderately well. 

\section{Discussion}

The linear spectral model we present here confirms what was predicted from earlier modelling of the optical-NIR colors of small TNOs. That is, the broad spectral behaviour of small TNOs is correlated with an object's optical slope. More specifically, the band-areas of most of the spectral features observable with JWST are well reproduced with the simple knowledge of an object's optical color and knowledge of its compositional class: either \bowl~ or \dd/\cliff.

Moreover, the trends in band-area with $o$ are moderately well described by linear modelling. Both of these properties must be true for the linear modelling of \citet{Fraser2012} and \citet{Fraser2023} to have any true compositional validity as both of these properties are predictions of the mixture models used to explain the distribution of observed TNOs colors they discuss.

The obvious interpretation of this result is that certain key materials largely govern the overall spectral shapes of a given spectral class, including their optical colors. The \org~ bodies show a clear trend of increasing organic absorption and decreasing CO and CO$_2$ absorption with increase in optical slope. The aliphatic materials are a red optical colorant and the CO and CO$_2$ ices are neutral reflectors. When mixed, these two materials could account for the trend of their relative absorption band areas with optical slope. The \bowls~ appear to broadly share the same type of optical colorant, but the neutral reflector is predominantly water-ice, though optical color increases with H$_2$O band area. \edit{Naively, this trend could be interpreted that redder objects possess greater abundances of surface water-ice. It is also possible however,} that variations in aliphatic content or grain size variations are enough to determine optical color. Alternatively, the trend with water-ice and $o$ is compatible with the suggestion by \citet{Grundy2009} that increasing concentrations of transparent ice in a mixture of ice and reddening agent could make that mixture redder.

Importantly, the specific compositional ratios of each type's key materials are not random, but must span a very limited  range of mixtures which are typical of a class, and those ratios are predictable by an object's optical color. For example, the bluest members of the \bowls~ will have a specific makeup of water-ice, OH, and CO$_2$ surface content, and the relative concentration of those materials and their grain sizes will follow a specific and predictable trend as ever redder \bowls~ are considered. Considering the end-members, the bluest \bowls~ will have the weakest apparent OH, H$_2$O and CO absorptions and the strongest CO$_2$ absorption, and the opposite for the reddest. Importantly, the specific ratios of those four components will be limited and predictable for the majority of \bowls.  Similar statements can be made about H$_2$O, CO$_2$, CO and the aliphatic organics which appear to be the key materials that govern the spectral behaviour of \org~ bodies, and will only commonly be present in a tight range of compositional mixtures that correlates with the optical colors of those objects. Importantly, this modelling also suggests that \cliffs~ and \dds~ are cosmogonically related. \edit{It is likely that the specific compositional ratios of the \org~ and \bowls~ will provide valuable information about the compositional makeup of the disc from whence these objects accreted. Significant additional work is required on this front.}

While the linear model captures a very broad range of spectral behaviour for a class, clearly there is a larger spectral diversity than can be captured with just a single parameter; the optical color. Equally as interesting as identifying the key materials that govern the spectra of a class of objects is identifying the materials that do not. For example, methanol is detected on many \org~ bodies, and shows a relation of band area and $o$ that is not accounted for by our linear modelling. This suggests that though methanol is common on those bodies, it seems that methanol is not a major driver of the overall chemistry that resulted in their broad spectral shapes of the \org~ bodies. It will be interesting to see what role methanol plays in the chemical evolution of these bodies. Similarly, while the linear model does a reasonable job of reproducing the $4.2\,\micron$ CO$_2$ feature, that material clearly shows a broader spectral diversity in the $2.7\,\micron$ shoulder feature than is accounted for by the simple linear model, showing a bimodality with optical color rather than a nearly linear trend. It remains to be seen what that source of that  bimodality is, though we speculate that size may play an as yet unrecognized factor in the \org~ objects. 

We have demonstrated that optical color is a first-order predictor of an object's specific compositional mixture. This finding enhances the diagnostic utility of spectro-photometric surveys. That is, one can provide moderately reliable assignations of an object's spectral type (\bowl~ or \org), rare outliers such as the blue binaries \citep{Wong2025b} or the Haumea family members \citep{Pinilla-Alonso2025} notwithstanding. The combination of orbital information with spectro-photometric taxonomic classification has provided useful leverage in determining the mass-density in the protoplanetesimal disc and the compositional distribution of TNOs therein prior to the dispersal of the disc \citep{Nesvorny2020,Marsset2023,Buchanan2026}. Unfortunately, the details inferred by these studies have been limited by unavailable compositional information. Spectral studies have used the presence of volatiles on some spectral classes and not others to infer specific heliocentric distances at formation of those bodies \citep{Brunetto2025}, but due to the lack of calibration of discovery bias, and overall limited sample size, spectra provide only limited information about the primordial mass and orbital distributions. Linking composition to optical color will greatly increase the potency of the massive sample of $\sim35,000$ TNOs that will have orbits and optical colors measured by  the pending Legacy Survey of Space and Time (LSST) from the Vera C. Rubin Telescope \citep{Kurlander2025}. That survey will provide high quality (g-r) and (r-z) measurements for a massive sample of objects, sufficient to provide taxonomic classification into BrightIR/NIRB and FaintIR/NIRF groups. The model here will then provide specific compositional inference for that massive sample. Focused effort will still be needed to gather detailed spectra to test if the known compositional trends with $o$ hold for a larger sample spanning a broader range of orbits and sizes. Those efforts combined with a link to the LSST sample should enable a significant improvement in our knowledge of the primordial mass and composition distributions of TNOs.

When a much larger sample is available, further work should include a search for what other parameters may act as predictors for the diversity that optical color does not predict. We speculate that optical albedo will be revealed as an important parameter. While no clear trend of albedo and spectral shape has emerged \citep{Pinilla-Alonso2025}, it seems plausible that the optical albedo of an object would be influenced by highly optically reflective ices. It seems plausible that the albedos of \cliffs~ and \dds~ will clump into two albedo groups much like they do in the $2.7\,\micron$ CO$_2$ band-area. Additional further work should be dedicated to revealing the specific compositional mixtures of the key materials as a function of optical color and to determining the end-member mixtures of each main compositional type. It is likely that those specific mixtures will be cosmogonically informative.

\clearpage

\bibliography{export-bibtex.bib}{}
\bibliographystyle{aasjournalv7}

\end{document}